\documentclass[prl,superscriptaddress,showkeys,showpacs,10pt,tightenlines,floatfix,twocolumn]{revtex4}
\usepackage[utf8]{inputenc}
\usepackage{amsmath}
\usepackage{amsfonts}
\usepackage{amssymb}
\usepackage{graphicx}
\usepackage{color}
\usepackage{caption}
\usepackage[list=true]{subcaption}

\newcommand{\mH}{\mathcal{H}}

\newcommand{\Gammam}{\Gamma^-}
\newcommand{\Gammap}{\Gamma^+}
\newcommand{\Gm}{G^-}
\newcommand{\Gp}{G^+}
\newcommand{\ta}{\tilde{a}}
\newcommand{\tL}{\tilde{L}}
\newcommand{\tLo}{\tilde{L}_0}

\newcommand{\sech}{\mathrm{sech}}

\graphicspath{{./images/},{./images/old/}}
	
\begin{document}
	\title{Continuous limit of a discrete stochastic model of cell migration}

\author{Nino Despeignes}
\author{Marc Durand}
\affiliation{Université Paris Cité, CNRS, Matière et systèmes complexes, F-75013 Paris, France}

\email{marc.durand@univ-paris-diderot.fr} 
\date{\today}

\begin{abstract}
We analytically derive the continuous limit of the Cellular Potts Model (CPM) for a one-dimensional cell subjected to constant and run-and-tumble driving forces. By coarse-graining the discrete lattice dynamics, we obtain the Fokker-Planck equations governing the cell's size and center-of-mass position. We show that in the low-force regime, the cell dynamics are accurately described by an overdamped Langevin equation. Beyond this regime, we expose intrinsic algorithmic artifacts, including a force-dependent diffusion coefficient, a non-linear force-velocity relationship, and the breakdown of the Einstein relation. We demonstrate that replacing the conventional Metropolis update rule with Glauber dynamics significantly mitigates these artifacts, broadening the physically valid parameter space. Our exact results bridge the gap between lattice-based simulations and continuous active matter models.
\end{abstract}
\maketitle

A key challenge in modeling the collective motion of multicellular systems lies in the need to treat cells as spatially extended objects that move, grow, divide, die, deform, and undergo neighbor exchanges. All these processes contribute to the macroscopic properties of a cell assembly. Various numerical models incorporating these processes have been developed using diverse approaches \cite{Osborne2017,Beatrici2023} and are typically employed by distinct communities. Among these, the Cellular Potts Model (CPM) is particularly popular within the computational biology community. Its numerous applications include biological morphogenesis, angiogenesis, vasculogenesis, tumor invasion, and wound healing. This model offers several advantages: its core principle (Monte Carlo dynamics) is straightforward to implement; cellular domains are discretized on a lattice, making topological changes in the cellular pattern (division, death, intercalation) extremely easy to handle; and cells maintain realistic, convex shapes \cite{Joanny2020}. Furthermore, it handles confluent and non-confluent cells, as well as tissues with free boundaries (e.g., wound healing), with equal ease.

However, like any computational model, the CPM also has some drawbacks. In particular, its dynamics arise from a stochastic optimization process (specifically, the minimization of a free energy functional) rather than being governed by a Langevin-type equation of motion that explicitly relates forces to motion. Consequently, the active matter community, interested in the fundamental aspects of collective cell motion, generally prefers vertex or Voronoi models \cite{Honda1980,Weliky1990,Fletcher2014,Barton2017,Rozman2024}. In those frameworks, cells assume polygonal (2D) or polyhedral (3D) shapes, and each vertex (or cell center) follows Langevin dynamics where vertex motility and diffusivity are simulation inputs rather than outputs. Note, however, that for both models (CPM and vertex/Voronoi), the derivation of the dynamics at the cell scale from a coarse-graining of the dynamics implemented at the lattice site or vertex/cell center scale remains an open problem.

The aim of this Letter is to partially address this question for the CPM. We first focus on the simple situation of a single cell evolving in a one-dimensional space and subjected to a constant force (due, for instance, to a uniform chemotactic gradient or constant migrating drive), and subsequently on a cell exhibiting run-and-tumble dynamics of its center of mass. Given the stochastic nature of the model, we derive the Fokker-Planck equations that govern the evolution of the cell size and position in the continuous limit, and establish analytic expressions for the cell's diffusion coefficient and mobility as functions of the simulation parameters. Two different Monte Carlo dynamics are investigated and compared for the lattice site update rule: Metropolis and Glauber dynamics. Our analysis also yields bounds on the simulation parameters, within which the cell dynamics are accurately described by a Fokker-Planck or Langevin equation. Several unexpected behaviors, hereafter termed artifacts, are highlighted when the parameters fall outside these bounds. However, we demonstrate that some of these artifacts are delayed when utilizing Glauber dynamics rather than Metropolis dynamics.

Several studies have empirically characterized the random walk \cite{Mombach1996,Guisoni2018,Guisoni2020} and drift velocity \cite{Marée2007} of a cell. There have been few attempts to derive the continuous limit of the CPM. In \cite{Turner2004}, the diffusion coefficient $D$ for a collection of noninteracting randomly moving cells was formally calculated from a one-dimensional Cellular Potts Model. However, $D$ was expressed as a sum rather than an analytic function dependent on the parameters of the CPM. In a series of papers \cite{Alber2006,Alber2007,Alber2007b,Alber2008}, Alber and coworkers studied the continuous limit of the CPM describing individual cell motion in a medium and in the presence of an external field with contact cell-cell interactions at low and high cell densities. However, the derived expression of $D$ in the absence of coupling with the field does not depend on the CPM parameters, in contradiction with \cite{Turner2004} and our own findings.

\paragraph{Cellular Potts Model.}
The Cellular Potts Model (CPM) represents biological cells as connected domains of identical indices $\sigma_i$ on a lattice, evolving through the stochastic minimization of a Hamiltonian $\mathcal{H}$ that typically accounts for volume conservation and interfacial adhesion. In a one-dimensional setting, the lattice reduces to a linear chain of sites $i$. The dynamics follow a Metropolis-like algorithm: at each elementary step, a target site $i$ and a neighboring source site $j$ are randomly selected, and the index update $\sigma_i \to \sigma_j$ is proposed. This change is accepted with a probability $P= \min\left(1, e^{-\beta\Delta \mathcal{H}}\right)$, where $\Delta \mathcal{H}$ is the associated change in energy, and $\beta^{-1}$ is an effective temperature which reflects the amplitude of membrane fluctuations, and is thus associated with the activity of the cell cortex.

The minimal Hamiltonian $\mathcal{H}$ contains a compression term for each cell, along with a boundary energy term related to cell-cell adhesion. However, in the one-dimensional case of interest here, the length of the boundary between two adjacent cells is constant, and cells never change neighbors; this constant term can therefore be discarded. Cell migration is modeled by adding a force term to the Hamiltonian, which then reads:
\begin{equation}
	\mathcal{H} = \sum_{i=1}^N \frac{\kappa}{2} (L_i - L_0)^2 -  \sum_{i=1}^N F_{i} \cdot X_i,
	\label{Hamiltonian}
\end{equation}
where the first term represents the elastic energy, with $L_i$ being the instantaneous length of cell $i$, $L_0$ its nominal length, and $\kappa$ the compressibility modulus. The second term represents the potential energy due to a driving force $\mathbf{F}_{i}$ (such as a chemotactic drive or a constant motile force) acting on the center of mass $X_i$ of cell $i$. This force can be constant over time, or it can evolve according to its own stochastic dynamics \cite{Kabla2012,Chiang2016,Guisoni2018}. Let $a$ denote the lattice site size. All lengths are therefore multiples of $a$, which in practice is set to $1$ but is retained here to keep track of its dimension. The coordinates $L_i$ and $X_i$ are linearly related to the vertex positions $x_i$ and $x_{i+1}$: $L_i=x_{i+1}-x_{i}$, $X_i=(x_{i}+x_{i+1})/2$. Note that confining cells to a one-dimensional space is not merely a theoretical abstraction; it corresponds precisely to experimental conditions realized \textit{in vitro} \cite{Jain2020}.

\begin{figure}
	\centering
	\includegraphics[width=\linewidth]{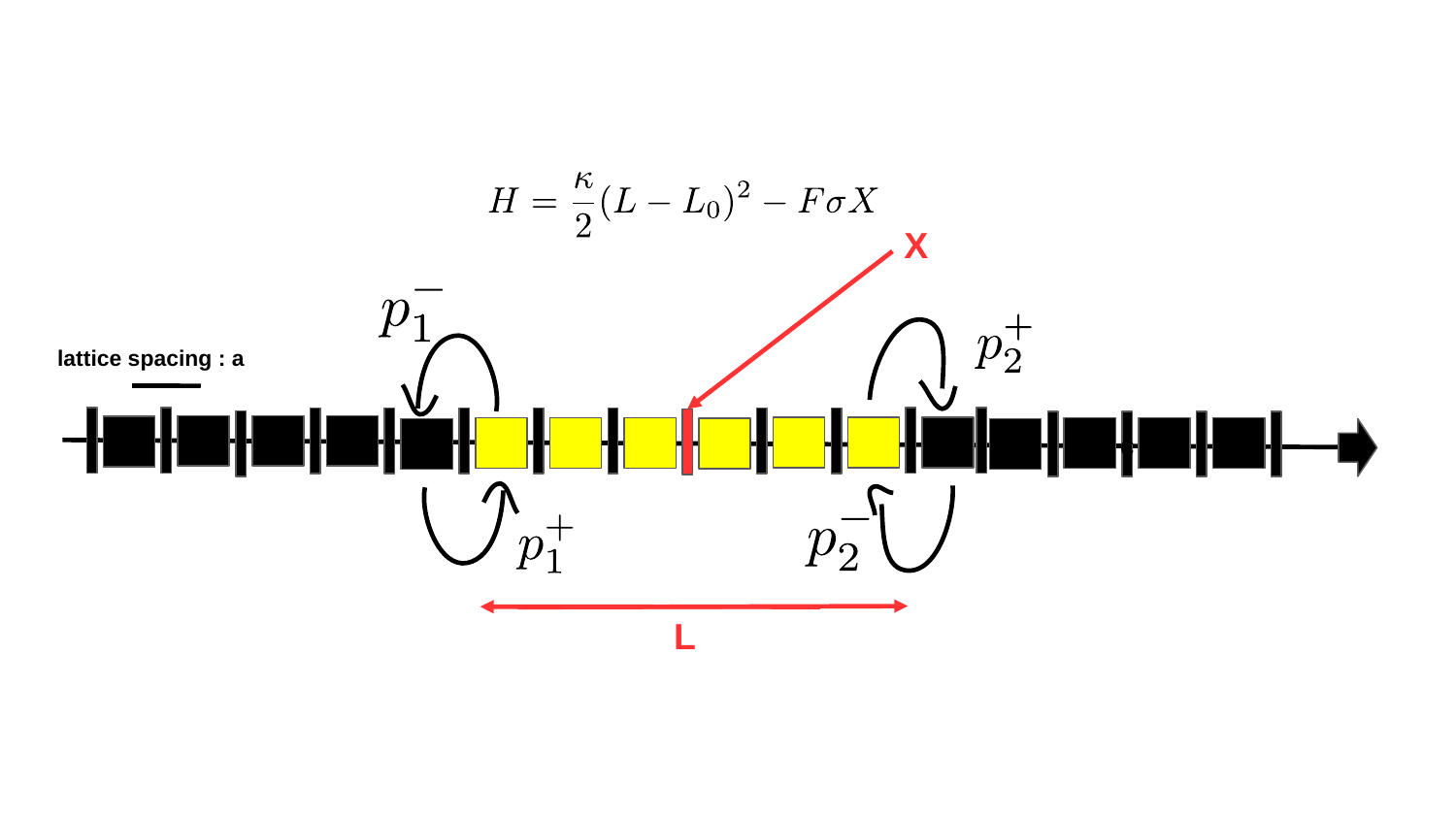}
	\caption{1D cell in the CPM model.}
	\label{fig:img1950}
\end{figure}

\paragraph{One cell under constant driving force}
We first study the case of a single cell surrounded by medium and subjected to a constant force $F$. Each time the position of one of its ends, denoted $x_1$ and $x_2$, is modified, its length $L$ changes by $\pm a$, and the position of its center of mass $X$ changes by $\pm a/2$. A master equation can be written for the positions of the ends, or equivalently for the variables $L$ and $X$. Using the latter allows for a more natural introduction of two distinct length and time scales. The master equation for $L$ reads:
 \begin{align}
	\partial_t P(L,t) \tau = & P(L+a,t) \Gammam(L+a)+P(L-a,t)\Gammap(L-a) \nonumber\\
	& -P(L,t) \left( \Gammam(L)+\Gammap(L) \right),
	\label{masterL}
 \end{align}
where $\tau$ is the time interval between two copy attempts ($\tau=1$ if time is measured in Monte Carlo steps), and $\Gamma^{\pm}(L)$ represents the probability that the cell length transitions from $L$ to $L\pm a$. The variation in length can arise from the displacement of either end, regardless of the cell's position, which is expressed as: $\Gamma^{\pm}(L)=\sum_{\{X\}}P(X,t|L)(p_1^\mp(X,L)+p_2^\pm(X,L))$, where $p_i^+(X,L)$ represents the probability that end $i$ ($i=1,2$) of the cell centered at $X$ with length $L$ takes a step to the right, and $p_i^-(X,L)$ the probability it takes a step to the left. Under Metropolis dynamics, we have $p_i^\pm = g\min \left(1,\exp (-\beta \Delta \mH_i^\pm)\right)$, where $g$ is the uniform selection probability of the target state, and considering the Hamiltonian expression (Eq. \ref{Hamiltonian} with $N=1$): $\Delta \mH_1^\pm = \kappa a \left( a/2 \mp (L-L_0)\right) \mp Fa/2$ and $\Delta \mH_2^\pm = \kappa a \left( a/2 \pm (L-L_0)\right) \mp Fa/2$. Note that the transition rates $p_i^\pm$ do not depend on $X$, yielding simply $\Gamma^{\pm}(L)= p_1^\mp(L)+p_2^\pm(L)$.

The Taylor expansion of the right-hand side of the master equation (\ref{masterL}) assumes that the transition rates $\Gamma^\pm$ vary slowly with $L$ \cite{VanKampen}, which requires $\beta \kappa a$ to be small. Specifically, taking $\vert L-L_0\vert \simeq \sigma_L$, where $\sigma_L$ is the standard deviation of the stationary distribution $P_s(L)$ to be determined \textit{a posteriori}, the criterion justifying the expansion is $\beta \kappa \sigma_L a \ll 1$. It is natural to introduce the dimensionless variables $\ta=\sqrt{\beta \kappa}  a$, $\tL = \sqrt{\beta \kappa}  L$, and $\tLo =\sqrt{\beta \kappa} L_0$. We also introduce the dimensionless force $f=\beta F a/2$. Expanding Eq. (\ref{masterL}) gives:
 \begin{align}
	\partial_t P(\tL,t)  = -\partial_{\tL} \left( v_{\tL} P(\tL,t) \right)+\partial^2_{\tL} \left( D_{\tL} P(\tL,t) \right)+\mathcal O(\ta^3),	
	\label{FPL}
\end{align}
with $v_{\tL}=\ta\left(\Gammap(\tL)-\Gammam(\tL) \right)/\tau$ and $D_{\tL}=\ta^2 \left(\Gammap(\tL)+\Gammam(\tL) \right)/(2\tau)$. To maintain consistency in the expansion to order $2$ in $\ta$ performed above, $\Gammap(\tL)-\Gammam(\tL)$ must be expanded to order $1$ and $\Gammap(\tL)+\Gammam(\tL)$ to order $0$. Because the transition rates $p_i^\pm$ are defined piecewise under Metropolis dynamics (depending on the sign of $\Delta \mH_i^\pm$), the expressions for $v_{\tL}$ and $D_{\tL}$ are defined over three distinct intervals of $\tL-\tLo$. Furthermore, the boundaries of these intervals differ depending on whether $F$ is greater or less than $B a$. We detail below the calculations for Glauber dynamics, which is commonly used in active matter modeling, and defer the analogous, albeit lengthier, calculations corresponding to Metropolis dynamics to the Supplemental Material \cite{SI}. Beyond simplifying the calculations, we will show that Glauber dynamics mitigates the artifacts observed with Metropolis.

The transition rates are defined by $p_i^\pm = g/ \left(1 + \exp (\beta \Delta \mH_i^\pm)\right)$, where the $\Delta \mH_i^\pm$ terms are identical to those in Metropolis. These rates thus remain independent of $X$, and one straightforwardly obtains \cite{SI} $v_{\tL}=-g\ta^2 (\tL-\tLo)~\sech^2 (f/2)/\tau+\mathcal{O}(\ta^4)$ and $D_{\tL}=g\ta^2/\tau+\mathcal{O}(\ta^4)$. Equation \ref{FPL} becomes:
\begin{align}
	\partial_t P(\tL,t)  & = \dfrac{g\ta^2}{\tau}\left[ ~\sech^2 \left(\dfrac{f}{2}\right)\partial_{\tL} \left(  (\tL-\tLo) P(\tL,t) \right) \right. \nonumber\\
	& \left.
+\partial^2_{\tL}  P(\tL,t) \right]+\mathcal O(\ta^4).
	\label{FPL2}
\end{align}
Note that the next term in the Taylor expansion of the master equation is $a^2\partial_{\tL} \left( v_{\tL} P(\tL,t) \right)/3! \sim \mathcal O(\ta^4)$. Thus, there is no third-order term in $\ta$. The solution to Eq. \ref{FPL2} is well known (it is the Fokker-Planck equation of a particle diffusing in a harmonic potential), and it can be shown \cite{SI} that it evolves over the characteristic time $T_{eq} \simeq (\sigma_L(f)/a)^2\tau / g$ toward a stationary normal distribution with zero mean and variance $\sigma_L^2(f)=1/(\beta  \bar\kappa(f))$, with $\bar\kappa(f)=\kappa~ \sech^2(f/2)$. The stationary distribution is therefore impacted by the presence of the force, which effectively decreases the cell's compression modulus $\bar\kappa(f)$.

Intriguingly, although the CPM dynamics satisfy detailed balance, the distribution converges to one that differs from the Gibbs-Boltzmann equilibrium distribution $P_{GB}(L) \propto \exp\left(-\beta \kappa(L-L_0)^2/2\right)$. This peculiarity stems from the fact that the CPM dynamics, wherein at most one end is displaced per time step, couple the evolutions of $L$ and $X$. We demonstrate in the Supplemental Material \cite{SI} that if a dynamics is chosen where $L$ and $X$ are modified independently, the distribution indeed converges to the Gibbs-Boltzmann equilibrium distribution for the variable $L$.

Note that the expression for the variance $\sigma_L^2(f)$ obtained above is only valid in the limit where the second-order Taylor expansion (Eq. \ref{FPL}) of the master equation is justified. However, the ratio of the next non-zero term in the expansion ($\mathcal O (\ta^4)$) to the first two terms ($\mathcal O (\ta^2)$) is proportional to $\ta^2\cosh^2\left(f/2\right)$ \cite{SI}. It is thus apparent that the permissible domains for the two dimensionless variables $\ta$ and $f$ are coupled: the smaller the value of one, the higher the other can be chosen. Note that this perfectly matches the condition $(\beta \kappa \sigma_L a)^2 \ll 1$ intuited when deriving Eq. \ref{FPL}. Furthermore, deriving this expression also required treating $L$ as a continuous, real variable by pushing the integration to $L\to -\infty$ for the normalization of $P_{s}(L)$. Ultimately, all these conditions require $\ta \ll \tilde \sigma_L(f) \ll \min (\ta^{-1}, \tLo)$, with $\tilde \sigma_L(f) = \sqrt{\beta \kappa} \sigma_L(f) = \cosh(f/2)$ for Glauber dynamics. Note that when the upper bound is limited by $\ta^{-1}$, both bounds (lower and upper) imply that $\ta^2 \ll 1$.

We now turn to characterizing the dynamics of the cell's center of mass $X$. The corresponding master equation reads:
 \begin{align}
	\partial_t P(X,t) \tau = & P(X+\dfrac{a}{2},t) \Gm+P(X-\dfrac{a}{2},t)\Gp \nonumber\\
	& -P(X,t) \left( \Gm+\Gp \right),
	\label{masterX}
\end{align}
where we have introduced the transition rates $G^{\pm}=\sum_{\{L\}}P(L,t|X)(p_1^\pm(L)+p_2^\pm(L))$. Since the rates $p_i^\pm$ are independent of $X$, we assume $P(L,t|X)=P(L,t)$. Moreover, we assume the simulation duration is much greater than the equilibration time $T_{eq}$, estimated previously, for the $L$ distribution, such that $P(L,t)=P_s(L)=\mathcal N(0,\sigma_L^2(F))$. The second-order expansion of Eq. \ref{masterX} yields:
 \begin{align}
	\partial_t P(X,t)  = - V_{}\partial_{X}  P(X,t)+D_{} \partial^2_{X} P(X,t)+\mathcal O(a^3),	
	\label{FPX}
\end{align}
where $V=a(\Gp-\Gm)/(2\tau)$ and $D=a^2(\Gp+\Gm)/(8\tau)$ correspond to the drift velocity and the diffusion coefficient of the cell center, respectively. After integrating over $L$, we obtain:
\begin{align}
	V_G &=  \dfrac{ g a}{\tau} \left(1-\dfrac{\ta^2}{4}\right) \tanh \left(\dfrac{f}{2} \right) \label{v-Glauber}\\
	D_G &= \dfrac{ g a^2}{4\tau}\left(1-\dfrac{\ta^2}{4}~\sech^2 \left(\dfrac{f}{2} \right)\right)
	\label{D-Glauber}
\end{align}
Deriving the expressions for $V$ and $D$ under Metropolis dynamics is more complex and is provided in the Supplemental Material. We just give here  the expressions in two limiting cases: first when $f \ll 1$: then
\begin{align}
		V_M &=\dfrac{g a}{\tau} \left( 1-\dfrac{\ta}{\sqrt{2\pi}}\right) f \label{v-Metropolis} \\
	D_M &=\dfrac{g a^2}{2 \tau} \left(1 - \dfrac{\ta}{\sqrt{2\pi}} \right)\label{D-Metropolis}
\end{align}
Note that there is a factor of 2 between the values of $V$ and $D$ obtained with the two algorithms in the limit $f \ll 1$ and $\ta \to 0$.
Then in the limiting case $f\gg \ta^2$ (but still $f \lesssim \mathcal O(1)$ to satisfy the condition $\ta \tilde \sigma_{L}(f) \ll 1$), the expressions simplify as:
\begin{align}
	V_{M} &= \dfrac{g a}{\tau} \left[ 1 - e^{-f} + \dfrac{\ta^2}{4} \left( e^{-f} - 1 \right) \right]\\
	D_M &= \dfrac{g a^2}{4\tau} \left[ 1 + e^{-f} - \dfrac{\ta^2}{4} \left( e^{-f} - 1 \right) \right]
\end{align}

\begin{figure*}[htb]
	\centering
		\subcaptionbox{\label{Vdrift_vs_ta_2algos}}{\includegraphics[width=0.45\textwidth]{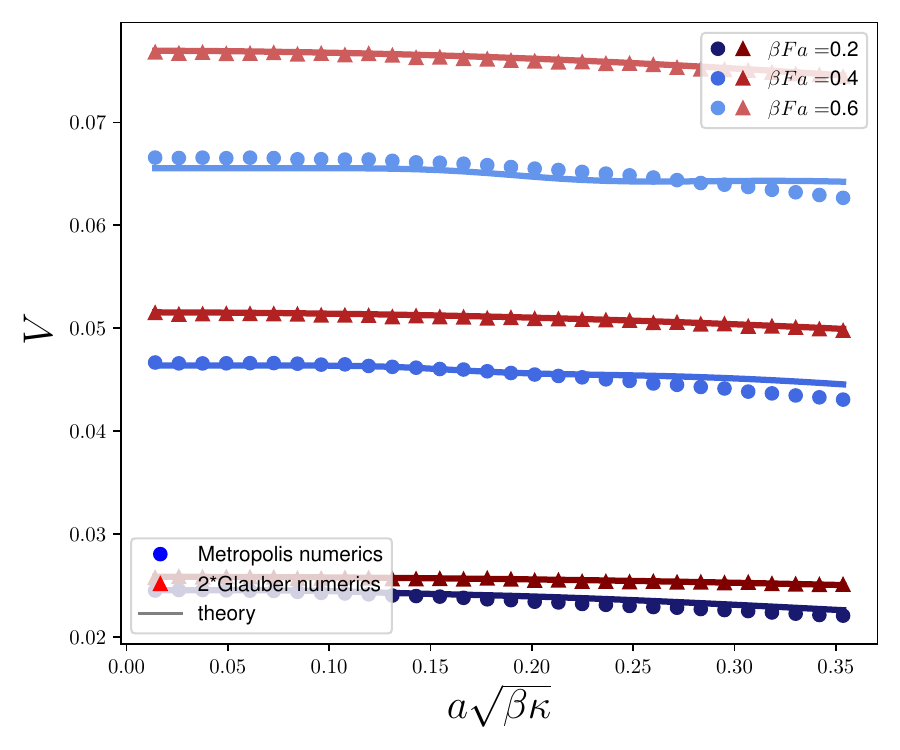}}
	\subcaptionbox{\label{Vdrift_vs_f_2algos}}{\includegraphics[width=0.45\textwidth]{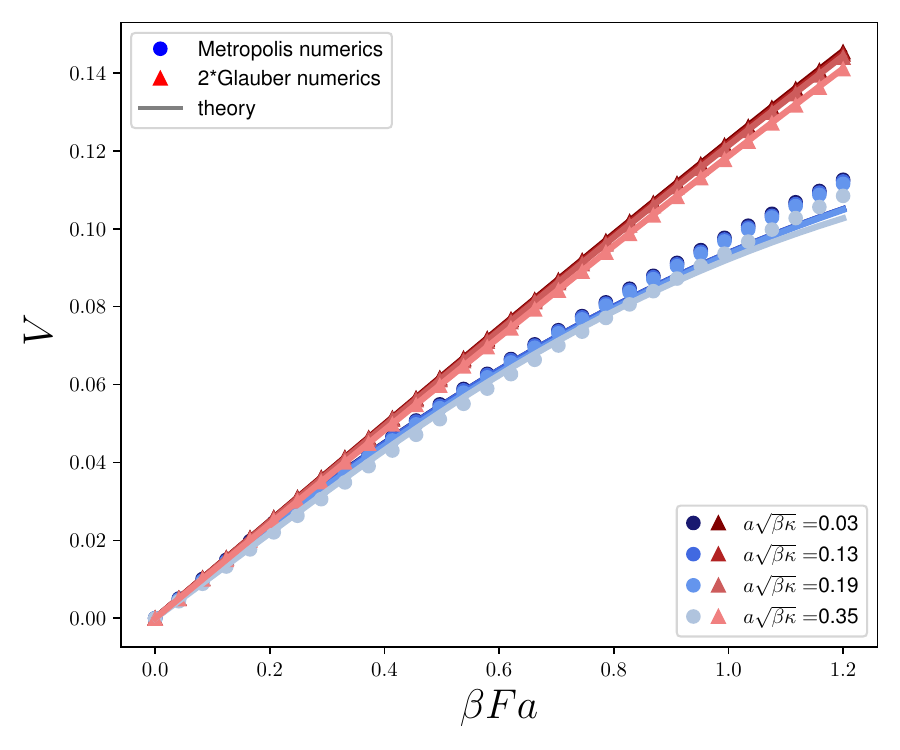}}
		\subcaptionbox{\label{fig:D_vs_ta_2algos}}{\includegraphics[width=0.45\textwidth]{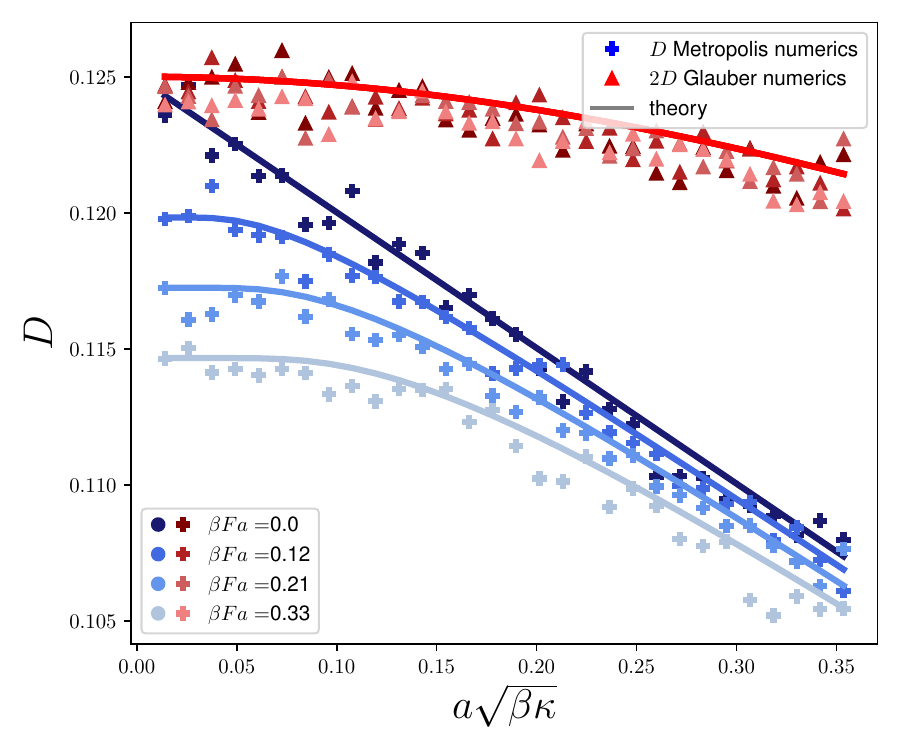}}
	\subcaptionbox{\label{fig:D_vs_f_2algos}}{\includegraphics[width=0.45\textwidth]{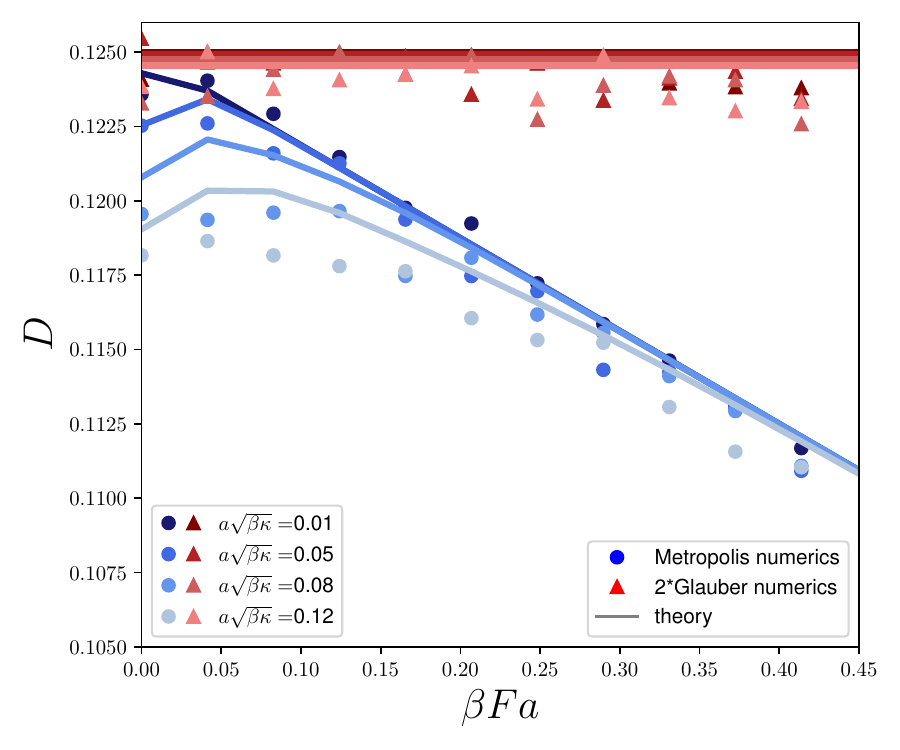}}
	\caption{Variation of the drift velocity end difffusive coefficient as a function of the two dimensionless parameters $\ta=\sqrt{\beta \kappa}a$ and $f=\beta F a/2$, for both the Metropolis (blue) and Glauber (red) dynamics.
\label{fig:Vdrift_et_D}}
\end{figure*}

Several remarks are in order regarding these expressions. First, the drift velocity is very weakly sensitive to the value of $\tau$ (and thus to the cell's compressibility), which is confirmed by simulations (see Figure \ref{Vdrift_vs_ta_2algos}). The dependence is however slightly more pronounced for Metropolis dynamics than for Glauber dynamics. Furthermore, the relationship between force and drift velocity is not linear, and the latter saturates as the force becomes very large. This phenomenon is well known \cite{Marée2007} and occurs because the center of mass displacement cannot exceed $a/2$ at each time step. However, it can be seen in Figure \ref{Vdrift_vs_f_2algos} that this non-linearity is shifted to much larger values of $F$ with Glauber dynamics, and the relationship is quasi-linear over practically the entire range of validity of the derived expressions ($f \lesssim 1$).

More surprisingly, the diffusion coefficient also depends on the force. Such a dependence does not exist when the drift velocity and diffusion coefficient are directly introduced into the stochastic dynamics of the vertices or the cell center of mass, as is customary in vertex/Voronoi cellular models ($\dot X=V+\sqrt{2D} \xi(t)$, where $\xi(t)$ is a Gaussian white noise); this dependence, specific to the CPM algorithm, arises because the transition rates $p_i^\pm(L)$ associated with length variations depend on $F$. As seen in Figure \ref{fig:D_vs_f_2algos}, this dependence is highly significant with the Metropolis algorithm but much less pronounced with the Glauber algorithm, which unfortunately is less commonly used in the CPM. 
Similarly, the dependance of the diffusion coefficient with the cell compressibility is much less pronounced with the Glauber than the Metropolis algorithm, as seen in Figure Figure \ref{fig:D_vs_ta_2algos}. Note the good agreement  between the numerical data and the analytical expressions for both algorithms. All these results show that Glauber dynamics suffers much less from the artifacts inherent to the CPM and should be favored over Metropolis dynamics.
Simulation details for Figure \ref{fig:Vdrift_et_D} are provided in the Supplemental Material.

Another particularity of these expressions is that they do not satisfy the Einstein relation: whether with Glauber or Metropolis dynamics, $V/D\neq \beta F$. This seems counter-intuitive at first: since these dynamics satisfy detailed balance, in an equilibrium situation, the distribution must necessarily correspond to the Gibbs-Boltzmann distribution $P_{GB}(X) \propto \exp\left(\beta F X\right)$. Yet, solving Eq. \ref{FPX} yields $P_s(X) \propto \exp\left(V X/D\right) \neq \exp\left(\beta F X\right)$. The origin of this paradox lies in the fact that Eq. \ref{FPX} does not accurately approximates the master equation \ref{masterX}, except in the limit $\beta F a \ll 1$. Indeed, we show \cite{SI} that the subsequent terms in the expansion are negligible only in the limit $D/V \gg a$, which, given the expressions for $V$ and $D$, implies $\beta F a \ll 1$. It is within this limit that the Fokker-Planck equation \ref{FPX} correctly describes the cell dynamics, and it is also within this limit that the Einstein relation is satisfied. Note that this condition necessarily requires taking $\ta \ll 1$ to respect the validity conditions of the distribution $P_s(L)$. For higher values of $f=\beta F a/2$, the equilibrium distribution (the stationary solution of the master equation \ref{masterX}) is indeed the Gibbs-Boltzmann distribution, but the Einstein relation no longer holds.

The derived expressions for $D$ and $V$ nevertheless remain valid beyond this low-force regime \cite{SI}. Thus, one must operate in the limit $f\ll 1$ for the diffusion coefficient to be independent of $f$, the drift velocity to vary linearly with $f$, the distribution $P_s(L)$ to coincide with the Gibbs-Boltzmann equilibrium distribution, and the center of mass dynamics to be described by the Fokker-Planck equation \ref{FPX}---meaning it obeys the overdamped Langevin equation: $\dot X=\mu F+\sqrt{2D} \xi(t)$, where the mobility $\mu=V/F$ and $D$ are independent of $F$. Their expressions, however, vary depending on the chosen dynamics (Metropolis or Glauber). It can be observed that the expressions for $V$ and $D$ are independent of the cell size $L_0$, which is also confirmed by our simulations (see \cite{SI}). In a $d$-dimensional space, as the center of mass displacement averages the displacements of boundary sites, we can anticipate these quantities to be inversely proportional to $L_0^{d-1}$. Furthermore, the dependence of $V$ and $D$ on the temperature $\beta^{-1}$ is non-trivial, as they are increasing functions of $\beta F a$ but decreasing functions of $\beta \kappa a^2$ (see Figure \ref{fig:Vdrift_et_D}).

\begin{figure}
\centering
\includegraphics[width=\linewidth]{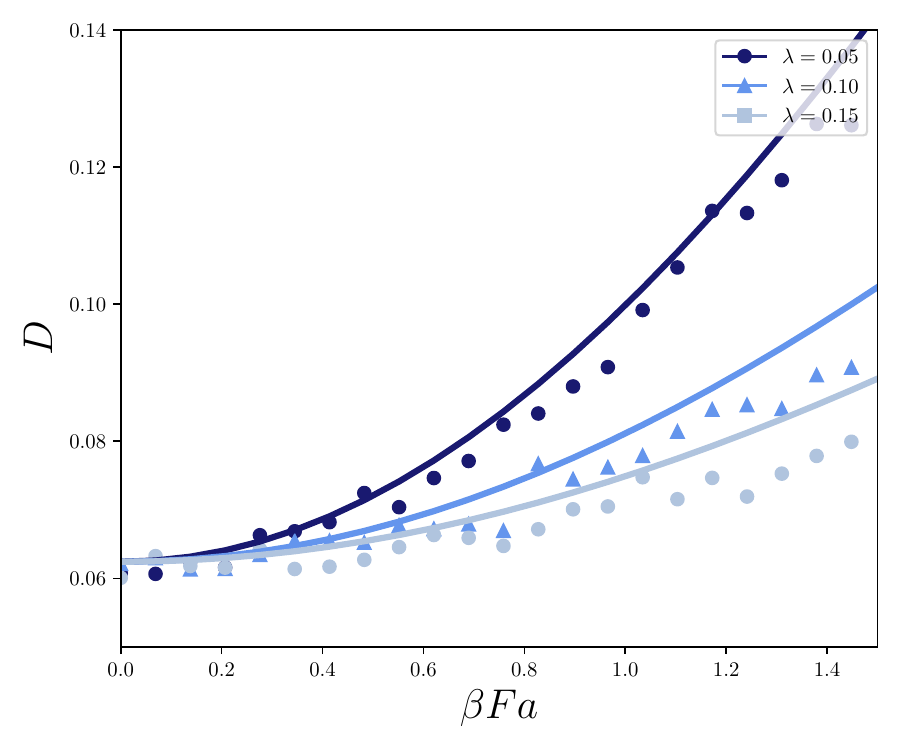}
\caption{Effective diffusive coefficient as a function of the adimensionalized force $\beta F a$.
}
\label{fig:active_diff}
\end{figure}

\medskip
\paragraph{One cell with run-and-tumble dynamics.} We now consider the case of a cell driven by a force exhibiting run-and-tumble dynamics, which in one dimension is equivalent to active Brownian dynamics \cite{Kabla2012,Chiang2016,Guisoni2018}. The force tumbles at a rate $\lambda$, and we denote by $F$ and $\epsilon$ ($\epsilon=\pm 1$) the magnitude and sign of this force, respectively (tumbling does not occur simultaneously with a vertex displacement). To study this scenario, we substitute $F \to \epsilon F$ in the Hamiltonian and the transition rate expressions. The cell state is defined by the three variables $L$, $X$, and $\epsilon$. It is easily verified that the transition rates satisfy $p_1^\pm (X,L,\epsilon) = p_2^\mp (X,L,-\epsilon)$. It follows that the evolution of the distributions $P(L, \epsilon,t)$ ($\epsilon=\pm 1$) is still governed by Eq. \ref{FPL}, meaning the stationary distribution remains unchanged ($P_s(L)\sim \mathcal N(0,\sigma_L^2(F))$). The evolution equation for $P(X,t)$, however, is modified to:
 \begin{align}
	\partial_t P(X,\epsilon,t)  & = -  v_{}\partial_{X}  P(X,\epsilon,t)+D_{} \partial^2_{X} P(X,\epsilon,t) \nonumber\\
&	+\lambda P(X,-\epsilon,t) - \lambda P(X,\epsilon,t),
	\label{FPXRT}
\end{align}
where $V$ and $D$ are the drift velocity and diffusion coefficient of a non-tumbling cell (Eqs. (\ref{v-Glauber})-(\ref{D-Glauber}) for a Glauber dynamics). We introduce $P(X,t)= P(X,+1,t)+P(X,-1,t)$ and $J =(P(X,+1,t)-P(X,-1,t))V$. Following the same procedure as in \cite{Solon2015} (see \cite{SI}), we find that at long times ($t\gg \lambda^{-1}\gg \tau$) and large spatial scales ($D \partial_X^2 J \ll \lambda J$), $P(X,t)$ obeys the diffusion equation $\partial_t P(X,t)   = D_\mathrm{eff} \partial^2_{X} P(X,t) $ with the effective diffusion coefficient $D_\mathrm{eff}=D + V^2/(2\lambda)$. Thus, at long times ($t\gg \lambda^{-1}$), there is no longer any drift, meaning the low-force regime ($\beta F a \ll 1$) is no longer a necessary condition for the evolution equation to be valid, contrary to the constant-force scenario ($\lambda \to 0$) studied earlier. Moreover, the diffusion coefficient is simply expressed as the sum of two terms: the diffusion coefficient induced by the cell fluctuations associated with the effective temperature $\beta^{-1}$, plus an active diffusion term that depends on the tumbling rate $\lambda$. However, this second term also depends on $\beta^{-1}$, preventing us from assigning each term to a specific source of activity (membrane activity vs. cell motility reorientation). We therefore suggest multiplying the force term by $\beta^{-1}$ in the Hamiltonian expression (Eq. \ref{Hamiltonian}) if one wishes to associate each term with a distinct source of activity. Figure \ref{fig:active_diff} confirms the dependence of $D_\mathrm{eff}-D$ on $V^2/\lambda$ predicted by our expression.

In conclusion, we have analytically characterized the CPM dynamics at the cellular scale, starting from the dynamics implemented at the lattice site scale, for a single cell in a 1D space. We derived the expressions for the drift velocity and diffusion coefficient when this cell is subjected to a constant driving force, and subsequently to a tumbling force. In the first case, we demonstrated that in a low-force regime ($\beta F a \ll 1$), the dynamics of the cell's position are described by an overdamped Langevin equation with a diffusion coefficient and drift velocity expressed in terms of the simulation parameters and the chosen lattice site update dynamics (e.g., Metropolis or Glauber). Beyond this low-force regime, the expressions for the diffusion coefficient and drift velocity remain derivable up to moderate forces, but the center of mass dynamics can no longer be described by a Langevin equation, and certain artifacts appear: effective cell softening and broadening of its length distribution, non-linear relationship between drift velocity and force, and force dependence of the constant diffusion coefficient. We highlighted that using Glauber dynamics instead of the Metropolis dynamics conventionally used in the CPM significantly mitigates some of these artifacts, and should thus be preferred. For a tumbling driving force, we established the expression for the new diffusion coefficient. At long times, it is expressed as the sum of two terms: the fluctuation-induced diffusion associated with the effective temperature $\beta^{-1}$, plus an active diffusion term depending on both $\beta^{-1}$ and the tumbling rate $\lambda$, preventing the identification of each term with a specific activity. Renormalizing the force $F$ by $\beta^{-1}$ in the Hamiltonian allows for this clear distinction. This work represents an important first step toward studying the coarse-grained dynamics of the Cellular Potts Model for 2D or 3D multicellular systems.

\bibliography{biblio}

\end{document}